\documentstyle[12pt]{article}
\newlength{\pubnumber} 

\catcode`\@=11
\def\section{\@startsection{section}{1}{\z@}{3.5ex plus 1ex minus .2ex}
 {2.3ex plus .2ex}{\large\bf}}
\def\subsection{\@startsection{subsection}{2}{\z@}{2.3ex plus .2ex}
 {2.3ex plus .2ex}{\bf}}

\begin{document}

\begin{titlepage}
\samepage{
\setcounter{page}{1}
\rightline{IASSNS-HEP-95/41}
\rightline{\tt hep-th/9505194}
\rightline{June 1995}
\vfill
\begin{center}
 {\Large \bf Spacetime Properties of (1,0) String Vacua\footnote{
       Talk presented at {\it Strings '95:  Future Perspectives
       in String Theory},
       held at the University of Southern California,
       Los Angeles, CA, 13--18 March 1995.  To appear in
       the Proceedings published by World Scientific.}\\}
\vfill
 {\large Keith R. Dienes\footnote{
   E-mail address: dienes@guinness.ias.edu.}\\}
\vspace{.12in}
 {\it  School of Natural Sciences, Institute for Advanced Study\\
  Olden Lane, Princeton, NJ~  08540~ USA\\}
\end{center}
\vfill
\begin{abstract}
  {\rm We discuss one of the generic spacetime consequences
  of having (1,0) worldsheet supersymmetry in tachyon-free string theory,
  namely the appearance of a ``misaligned supersymmetry''
  in the corresponding spacetime spectrum.  Misaligned supersymmetry
  is a universal property of (1,0) string vacua which
  describes how the arrangement of bosonic and fermionic states at
  all string energy levels conspires to preserve finite string amplitudes,
  even in the absence of full spacetime supersymmetry.
  Misaligned supersymmetry also constrains the degree to which spacetime
  supersymmetry can be broken without breaking modular invariance,
  and is responsible for the vanishing of various mass supertraces
  evaluated over the infinite string spectrum.  }
\end{abstract}
\vfill}
\end{titlepage}

\setcounter{footnote}{0}

\def\beq{\begin{equation}}
\def\eeq{\end{equation}}
\def\beqn{\begin{eqnarray}}
\def\eeqn{\end{eqnarray}}

\def\ie{{\it i.e.}}
\def\eg{{\it e.g.}}
\def\half{{\textstyle{1\over 2}}}
\def\third{{\textstyle {1\over3}}}
\def\quarter{{\textstyle {1\over4}}}

\def\Htw{{\tilde H}}
\def\chibar{{\overline{\chi}}}
\def\qbar{{\overline{q}}}
\def\ibar{{\overline{\imath}}}
\def\jbar{{\overline{\jmath}}}
\def\Hbar{{\overline{H}}}
\def\Qbar{{\overline{Q}}}
\def\abar{{\overline{a}}}
\def\alphabar{{\overline{\alpha}}}
\def\betabar{{\overline{\beta}}}
\def\tautwo{{ \tau_2 }}
\def\thetatwo{{ \vartheta_2 }}
\def\thetathree{{ \vartheta_3 }}
\def\thetafour{{ \vartheta_4 }}
\def\ttwo{{\vartheta_2}}
\def\tthree{{\vartheta_3}}
\def\tfour{{\vartheta_4}}
\def\ti{{\vartheta_i}}
\def\tj{{\vartheta_j}}
\def\tk{{\vartheta_k}}
\def\calF{{\cal F}}
\def\smallmatrix#1#2#3#4{{ {{#1}~{#2}\choose{#3}~{#4}} }}
\def\ab{{\alpha\beta}}
\def\Minv{{ (M^{-1}_\ab)_{ij} }}
\def\bone{{\bf 1}}
\def\ii{{(i)}}


\def\inbar{\,\vrule height1.5ex width.4pt depth0pt}

\def\IC{\relax\hbox{$\inbar\kern-.3em{\rm C}$}}
\def\IQ{\relax\hbox{$\inbar\kern-.3em{\rm Q}$}}
\def\IR{\relax{\rm I\kern-.18em R}}
 \font\cmss=cmss10 \font\cmsss=cmss10 at 7pt
\def\IZ{\relax\ifmmode\mathchoice
 {\hbox{\cmss Z\kern-.4em Z}}{\hbox{\cmss Z\kern-.4em Z}}
 {\lower.9pt\hbox{\cmsss Z\kern-.4em Z}}
 {\lower1.2pt\hbox{\cmsss Z\kern-.4em Z}}\else{\cmss Z\kern-.4em Z}\fi}

\hyphenation{su-per-sym-met-ric non-su-per-sym-met-ric}
\hyphenation{space-time-super-sym-met-ric}
\hyphenation{mod-u-lar mod-u-lar--in-var-i-ant}


\setcounter{footnote}{0}

In string theory, the interplay between worldsheet symmetries and their
consequences
in spacetime remains largely mysterious.  Certain results, however,
indicate strong connections between the two.  For example, it is well-known
that
$N=4$ supersymmetry on the worldsheet implies $N=2$ supersymmetry
in spacetime, and likewise it has been demonstrated that
$N=2$ supersymmetry on the worldsheet implies $N=1$ supersymmetry
in spacetime.  In this talk, we shall consider the more general
situation from which these results might follow as special cases.
In particular, we shall discuss some of the spacetime consequences
of $N=1$ worldsheet supersymmetry.

There are various reasons why this is an important question.
$N=1$ worldsheet supersymmetry is the defining property
of the superstring and heterotic string theories,
and it is in fact this feature which is single-handedly responsible
for the introduction of spacetime fermions into the resulting string spectrum.
It is therefore interesting to determine whether $N=1$ worldsheet supersymmetry
is also sufficiently powerful to constrain the {\it distribution}\/ of
these fermions relative to the bosons.
Clearly we do not expect exact boson/fermion degeneracies,
as occur in the more restrictive cases with spacetime supersymmetry
resulting from $N=2$ or $N=4$ worldsheet supersymmetry,
but we might expect that some more general constraints nevertheless
control their distribution.
Another reason for investigating this issue is to uncover
some of the hidden stringy mechanisms whereby super- or heterotic string
theories achieve finiteness even without spacetime supersymmetry.
For example, it is well-known that
the string one-loop vacuum energy (cosmological constant)
$\Lambda$ is a finite quantity
in these theories, {\it even without spacetime supersymmetry}\/;
in an ordinary non-supersymmetric field theory this quantity would diverge.
Clearly, the string theory differs from field theory in providing
infinite towers of massive (Planck-scale) states, and it is well-understood
how, through the requirement of modular invariance,
the presence of this tower succeeds in removing the ultraviolet divergences.
What is perhaps less clear, by contrast, is how the bosonic and fermionic
states ultimately conspire to arrange themselves level-by-level throughout this
tower in order to achieve this remarkable result.

Recently it has been shown \cite{mis} that the answers
to these questions involve
a hidden so-called ``misaligned supersymmetry'' which persists in the
string spectrum, even without full spacetime supersymmetry.
Indeed, this ``misaligned supersymmetry''
is a generic property of modular-invariant tachyon-free $(1,0)$
string vacua, and appears for {\it any}\/ spacetime dimension $D>2$ and
for {\it any}\/ compactification mechanism.
Furthermore, it has been shown that this misaligned supersymmetry
is the underlying symmetry responsible for the finiteness of the
cosmological constant in the absence full spacetime supersymmetry,
and in fact explains how
string finiteness is ultimately reconciled with the presence of
exponentially growing numbers of string states
throughout the infinite towers in the string spectrum.
Moreover, misaligned supersymmetry in principle also constrains
the degree to which spacetime supersymmetry may be broken
in string theory without destroying modular invariance and
the resulting finiteness of string amplitudes.
Indeed, misaligned supersymmetry is sufficiently powerful
to guarantee the vanishing of various mass supertraces ${\rm Str}\, M^n$
in string theory, even without spacetime supersymmetry \cite{supertraces}.
Thus, in some sense, misaligned supersymmetry lies at the root
of many of the remarkable properties that string theory exhibits.

In the remainder of this talk, we shall outline some spacetime
consequences
of misaligned supersymmetry.  Further discussion and details can
be found in Ref.~\cite{mis}.

We begin by considering how states are typically arranged in string theory.
In general, the string spectrum consists of a collection of infinite
towers of states:  each tower corresponds to a different {\it sector}\/
of the underlying string worldsheet theory, and consists of
a ground state with a certain vacuum energy $H_i$
and infinitely many higher excited states with worldsheet
energies $n=H_i+\ell$ where $\ell\in \IZ$.
The crucial observation, however, is that the different sectors in
the theory will in general be {\it misaligned}\/
relative to each other, and start out with
different vacuum energies $H_i$ (modulo 1).
For example, while one sector may contain states with integer energies $n$,
another sector may contain states with $n\in\IZ+1/2$, and another
contain states with $n\in \IZ+1/4$.
Thus each
sector essentially contributes a separate set of states to the
total string spectrum, and we can denote the net degeneracies of
these states
from the $i^{\rm th}$ individual sector as $\lbrace a_{nn}^{(i)}\rbrace$,
where $n\in \IZ + H_i$.
Thus, $a_{nn}^{(i)}$ represents the number of spacetime bosons minus
fermions in the $i^{\rm th}$ sector of the theory
having spacetime (mass)$^2= n\in \IZ+H_i$.

For each sector $i$, let us now take the next step and imagine analytically
continuing the set of numbers $\lbrace a_{nn}^{(i)}\rbrace$
to form a smooth function $\Phi^{(i)}(n)$  which not only reproduces $\lbrace
a_{nn}^{(i)}\rbrace$ for the appropriate values $n\in \IZ+H_i$,
but which is continuous as a function of $n$.
These functions $\Phi^{(i)}(n)$ clearly must not only
exhibit the leading Hagedorn exponential dependence $e^{C\sqrt{n}}$,
but must also contain all of the subleading behavior as well
so that exact results can be obtained
for the relevant values of $n$.
Such continuations are unique and well-defined, and
may be easily generated \cite{HRKV}.

Given that such functions $\Phi^{(i)}(n)$ exist, misaligned supersymmetry
is then characterized by the cancellation of the sum of these
functional forms over all sectors in the theory:
\beq
     \sum_{i} \, \Phi^{(i)}(n)~ = ~ 0~.
\label{sectoraverage}
\eeq
Note that this is a cancellation in
the {\it functional forms}\/ $\Phi^{(i)}(n)$,
and {\it not}\/ a cancellation in the actual numbers of
states $\lbrace a_{nn}^{(i)}\rbrace$.

In order to see the effect of this cancellation on the actual numbers of states
$\lbrace a_{nn}^{(i)}\rbrace$, let us examine
a simple hypothetical example, a toy string model containing only two sectors
$A$ and $B$.
For the sake of concreteness, let us assume that these two sectors
have different vacuum energies, with $H_A=0$ (modulo 1)
and $H_B=1/2$ (modulo 1).
We thus have two separate towers of states in this theory,
with degeneracies $\lbrace a_{nn}^{(A)}\rbrace$ situated at energy
levels $n\in \IZ$, and degeneracies $\lbrace a_{nn}^{(B)}\rbrace$ situated
at energy levels $n\in \IZ+1/2$ (in units of the Planck mass $M_0$).
Then if the corresponding functional forms that describe these degeneracies
are $\Phi^{(A)}(n)$ and $\Phi^{(B)}(n)$ respectively, then
misaligned supersymmetry implies that $ \Phi^{(B)}(n) =- \Phi^{(A)}(n)$.
However, due to the misalignment between the two sectors in this
hypothetical example, the actual value of each
individual $a_{nn}$ will be $\Phi^{(A)}(n)$ if $n\in \IZ$,
 {\it or}\/ $\Phi^{(B)}(n)$ if $n\in \IZ+1/2$.
This behavior is sketched in Fig.~1.
Thus, we see that misaligned supersymmetry leads to an
 {\it oscillation}\/ in which any given boson surplus at a certain
energy level implies a larger fermion surplus at a higher
energy level, followed by an even larger boson surplus at an even
higher level, and so forth throughout the string spectrum.

\input epsf.tex      
\begin{figure}[thb]
\centerline{\epsfxsize 4.0 truein \epsfbox {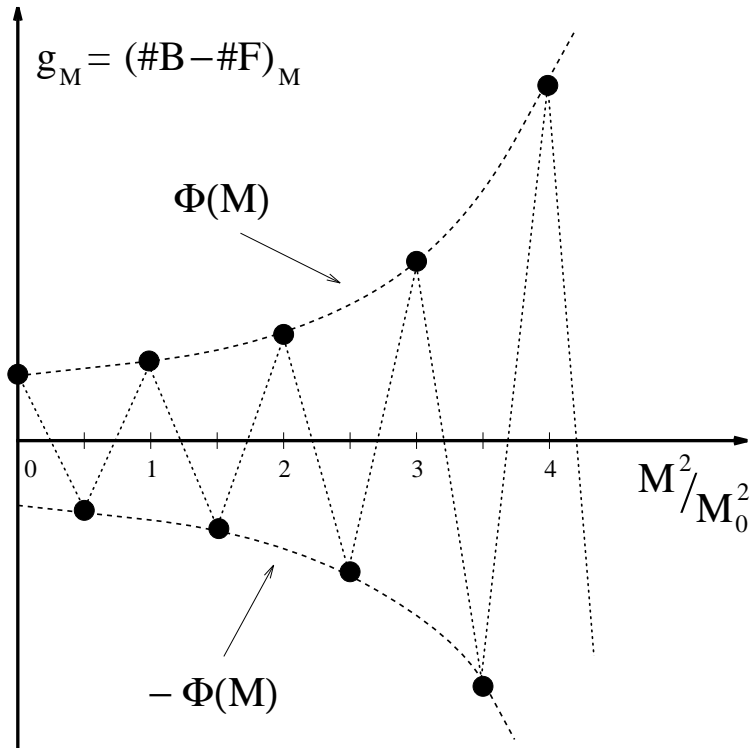}}
\caption{The net number of physical states
   $a_{nn}$ for the two-sector model
   discussed in the text, plotted versus energy $n$
   [equivalently the spacetime (mass)$^2$].}
\end{figure}

Such oscillatory behavior appears in any modular-invariant tachyon-free
theory {\it regardless}\/ of the number of sectors present, with
the simple sketch in Fig.~1 becoming more complicated for more complex
string models.  In any case, however, the cancellation
in Eq.~(\ref{sectoraverage}) is always preserved, with delicately balanced
boson/fermion oscillations persisting throughout the infinite string spectrum.

It is clear that spacetime supersymmetry is a special
case of this generic behavior,
for in this case we have $a_{nn}=0$ level-by-level,
and the ``amplitude'' of this oscillation is zero.
Thus, if spacetime supersymmetry is to be broken in such a way
that no physical tachyons are introduced and modular invariance is
be maintained (as required for a self-consistent string theory),
then we can at most ``misalign''
this bosonic and fermionic cancellation, introducing a mismatch
between the bosonic and fermionic state degeneracies at each level
in such a way that a carefully balanced ``misaligned supersymmetry'' survives.
It is interesting to see which classes of
physical super\-sym\-metry-break\-ing scenarios do not lead to
such behavior, and are thereby precluded.
For example, we can already rule out any scenario in which
the energies of bosonic and fermionic states are merely shifted
relative to each other by some amount $\Delta n$.
Instead, we would need to simultaneously
create a certain number $\Phi(n+\Delta n)-\Phi(n)$ of additional
states (presumably winding-mode states coming down from higher mass
levels) so that the cancellation
of the functional forms $\Phi$ is still preserved.
Such supersymmetry-breaking scenarios
are currently being investigated.

There are also a number of other potential applications and
extensions to misaligned supersymmetry.
For example, misaligned supersymmetry should be particularly relevant to
any system in which the asymptotic numbers
of high-energy states plays a crucial role, such as in
string thermodynamics  and the possible string phase transition.
In addition, we would also like to understand the role
that misaligned supersymmetry plays in ensuring finiteness
to {\it all}\/ orders (not just one-loop),
and also for all $n$-point functions.  Clearly, this requires
extending our results to include the {\it unphysical}\/ string states,
as well as string interactions.
Such work is in progress.


\bigskip
\bigskip
\leftline{\large\bf Acknowledgments}
\medskip

We thank the organizers of Strings '95 for the 4 minutes
and 40 seconds during which this talk was (quickly!)\ presented,
and thank R. Myers for comments on the manuscript.
This work was supported in part by DOE Grant No.\ DE-FG-0290ER40542.


\end{document}